\documentclass[
 aps, pra,
 amsmath,amssymb,
 11pt,
 final,
tightenlines,
 twoside,
 twocolumn,
 nofloats,
nofootinbib,
 superscriptaddress,
showkeys,
showkeywords,
 ]
{revtex4}
\usepackage{booktabs}
\usepackage{upgreek}
\usepackage{url}
\usepackage{natbib}
\usepackage{appendix}
\usepackage[utf8]{inputenc}
\usepackage[english]{babel}
\usepackage{graphicx}
\usepackage{dcolumn}
\usepackage{bm}
\usepackage{longtable}
\input{maik.rty}

\setcitestyle{authoryear,round}
\def\squareforqed{\hbox{\rlap{$\sqcap$}$\sqcup$}}

\def\sq{\ifmmode\squareforqed\else{\unskip\nobreak\hfil
\penalty50\hskip1em\null\nobreak\hfil\squareforqed
\parfillskip=0pt\finalhyphendemerits=0\endgraf}\fi}

\def\degr{\hbox{$^\circ$}}

\def\utw{\smash{\rlap{\lower5pt\hbox{$\sim$}}}}

\def\udtw{\smash{\rlap{\lower6pt\hbox{$\approx$}}}}

\def\diameter{{\ifmmode\mathchoice
{\ooalign{\hfil\hbox{$\displaystyle/$}\hfil\crcr
{\hbox{$\displaystyle\mathchar"20D$}}}}
{\ooalign{\hfil\hbox{$\textstyle/$}\hfil\crcr
{\hbox{$\textstyle\mathchar"20D$}}}}
{\ooalign{\hfil\hbox{$\scriptstyle/$}\hfil\crcr
{\hbox{$\scriptstyle\mathchar"20D$}}}}
{\ooalign{\hfil\hbox{$\scriptscriptstyle/$}\hfil\crcr
{\hbox{$\scriptscriptstyle\mathchar"20D$}}}}
\else{\ooalign{\hfil/\hfil\crcr\mathhexbox20D}}%
\fi}}

\newcommand{\apjl}{Astrophys.~J.}

\begin{document}

\selectlanguage{english}

\keywords{galaxies: groups: general---galaxies: haloes---galaxies: spiral}
\newcolumntype{.}{D{.}{.}{-1}}
\newcolumntype{0}{D{.}{.}{0}}


\title{Satellites around Edge-on Galaxies. I.~Dynamical Masses}

\author{\firstname{D.~V.}~\surname{Smirnov}}
\affiliation{Saint Petersburg State University, Saint Petersburg,
199034 Russia}

\author{\firstname{D.~I.}~\surname{Makarov}}
\affiliation{Special Astrophysical Observatory,  Russian Academy of Sciences,
Nizhnii Arkhyz, 369167 Russia}
\email{dim@sao.ru}

\author{\firstname{I.~D.}~\surname{Karachentsev}}
\affiliation{Special Astrophysical Observatory,  Russian Academy of Sciences,
Nizhnii Arkhyz, 369167 Russia}
\received{May 30, 2023} \revised{August 16, 2023} \accepted{August 18, 2023}

\begin{abstract}
We have undertaken a search for satellites around edge-on galaxies in the EGIPS catalog, which contains
16\,551~objects with declinations above~$-30\degr$. We searched
for systems with a central galaxy dominating in brightness by at
least $1^{\rm m}$ compared to its companions. As a result, we discovered
1097~candidate satellites around  764 EGIPS galaxies with
projected distances less than  500~kpc and a radial velocity
difference less than  300~km\,s$^{-1}$. Of these,  757~satellites
around 547~central galaxies have radial velocity
accuracies  higher than 20~km\,s$^{-1}${} and satisfy the
gravitationally bound condition. The ensemble of satellites is
characterized by an average projected distance of 84~kpc and an
average radial velocity dispersion of 103~km\,s$^{-1}${}. Treating
small satellites as test particles moving on isotropic orbits
around central EGIPS galaxies, we determined the projected
(orbital) masses of the edge-on galaxies. Within the luminosity
range of
 $1.3 \times 10^{10}$ to $42 \times 10^{10}$~$L_\odot$, the total
 mass of the systems is well described by a linear dependence
 $\log M_p \propto 0.88 \log \left\langle L_K \right\rangle_\mathrm{g}$
 with an average total mass-to-$K$-band luminosity equal
 to $(17.5 \pm 0.8)\,M_\odot/L_\odot$, which is typical for
  nearby spiral galaxies such as the Milky Way,  M\,31 and M\,81.
\end{abstract}

\maketitle

\section{INTRODUCTION}

According to the $\Lambda$CDM standard cosmological model, the
formation of galaxies in a diffuse medium takes place inside
potential wells formed by fluctuations in the distribution of dark
matter (Wechsler and
Tinker, 2018). Only about 5\% of the total
density of the Universe is taken up by visible baryonic matter of
galaxies (stars, gas and dust) (Aghanim et al., 2020).
Traditional methods of determining the total mass of a galaxy,
including the dark halo, from the rotation curve or from the stellar
radial velocity dispersion (see,
e.g., the survey by Zasov et al., 2017) are not sufficiently rigorous, since
they imply an extrapolation of observational data far beyond  the optical boundaries of the galaxy. Practically, two
methods are suitable for direct estimation of the total mass: an
analysis of the effects of weak gravitational lensing
(e.g., Viola et al., 2015), which also requires a
number of model assumptions, and an analysis of the kinematics of
satellites surrounding the galaxy. Both methods have their
advantages and shortcomings. The required accuracy of these
methods is reached by applying them to a large number of isolated
galaxies, selected by morphological type or some other
characteristics. The total masses of galaxies of various
categories were determined from the kinematics of their satellites
in may works: for spiral galaxies
(Zaritsky et al., 1993, 1997; Zaritsky and
White, 1994), for nearby galaxies of various types
in the Local Volume (Karachentsev and Kudrya, 2014; Karachentsev and Kashibadze,
2021), for especially
isolated galaxies (Karachentseva et al., 2021), and
for thin late-type spirals seen edge-on (Karachentsev
et al., 2016).

Karachentsev and Kudrya (2014) obtained a median estimate of the
total (projected) mass-to-K-band luminosity ratio of $31\,M_{\odot}/L_{\odot}$
considering the relative radial velocities and projected distances
of satellites around 15 nearby galaxies similar to the Milky Way
and Andromeda.
 Later Karachentsev and
Kashibadze (2021) used a richer sample of
298~satellites around 25~bright Milky Way-type galaxies in the
Local Volume and found for them an average ratio of
 \mbox{$\langle M_p/L_K \rangle = (31 \pm 6)\,M_{\odot}/L_{\odot}$}.
  Additionally, a higher $M_p/L_K$ value was noted for a
  sample of 47~low-brightness galaxies
 similar to the neighboring M\,33 galaxy and the Magellanic Clouds.
 The average $M_p/L_K$ ratio for the disk-dominated galaxies,
$(17.4 \pm 2.8)\,M_{\odot}/L_{\odot}$, turned out to be
significantly lower than for the bulge-dominated galaxies, \mbox{$(73
\pm 15)\,M_{\odot}/L_{\odot}$}.

Karachentseva et al. (2021) used a similar approach for estimating
the total masses of particularly isolated galaxies. Based on the
radial velocities and projected distances of 141~satellites, they
obtained  \mbox{$\langle M_p/L_K \rangle = (20.9 \pm
3.1)$~$M_{\odot}/L_{\odot}$}, noting the upward trend of this
ratio towards earlier-type galaxies.

An independent determination of $M_p/L_K$ for Sc--Sd galaxies was
carried out by Karachentsev
and Karachentseva (2019). An average ratio of
$(20\pm3)\,M_{\odot}/L_{\odot}$ was obtained from 43~satellites
for 220~galaxies without visible bulges and oriented face on,
which is in agreement with the previous estimates.

In this work we consider the kinematics of satellites that we
discovered around edge-on galaxies in the EGIPS
catalog (Makarov et al., 2022). Edge-on galaxies attract our
attention by the fact that such an orientation gives one an
opportunity to investigate directly the distribution of satellites
with respect to the disk plane of the galaxies, which is
impossible in other scenarios. The EGIPS catalog is the largest
sample of edge-on galaxies to date, which allows one to
significantly increase the statistics and, as a consequence,
enhance the reliability of the obtained results. In this work we
use the standard cosmological $\Lambda$CDM model with parameters
\mbox{$H_0=72.0$~km\,s$^{-1}$ Mpc$^{-1}$}, \mbox{$\Omega_m=0.3$,
$\Omega_\Lambda=0.7$.}

\section{SATELLITE SAMPLING}

The sample of central objects is based on the new catalog of edge-on galaxies
(EGIPS, Makarov et al., 2022), discovered in the images of the \linebreak
\mbox{Pan-STARRS1 DR2}(PS1, Chambers et al., 2016) survey. It contains data
on 16\,551~galaxies located above $\delta>-30^\circ$. 
The photometry that we carried out using the {\tt
SExtractor} (Bertin and Arnouts, 1996) package in all five
filters ($g$, $r$, $i$, $z$, $y$) of the PS1 survey provides an
unbiased flux estimate with an accuracy of about 0.05 for galaxies
fainter than $r\simeq13\,.\!\!^{\rm m}8$. Most EGIPS galaxies
(about 63\%) have redshift measurements with a median value of
$cz_\mathrm{CMD}\approx12\,000$~km\,s$^{-1}$, corresponding to a
depth of about 170~Mpc. To avoid uncertainties in the galaxy
distance estimates due to high peculiar velocities in the nearby
Universe, we restricted our study to EGIPS galaxies with
\mbox{$cz_\mathrm{LG}\geq 2000$~km\,s$^{-1}$.} Due to
observational selection the number of galaxies with known
redshifts falls drastically after \mbox{$cz_\mathrm{CMD} \approx
12\,000$~km\,s$^{-1}$}. Statistics are extremely sparse for systems
with redshifts above 30\,000~km\,s$^{-1}${}
 (there are about 10~systems with higher velocities), and
 we therefore limited the sample by a maximum redshift of
  $cz_\mathrm{CMD}=30\,000$~km\,s$^{-1}$.

The sample of candidate satellites was formed in four stages. The
first step consisted of compiling a list of approximately
3.2~million galaxies with known redshifts and photometry from
SDSS\,DR17 (Abdurro'uf et al., 2022) survey data and the
HyperLeda\footnote{\url{https://leda.univ-lyon1.fr/}}
(Makarov et al., 2014) database. The use of SDSS provides
highly homogeneous data; however, it presents a number of
difficulties. For a large number of bright galaxies
(\mbox{$m_r\lesssim 14\,.\!\!^{\rm m}5$}) that fall into the
photometric part of the SDSS survey, no radial velocity or
magnitude measurements were carried out. Additionally, there is
the well-known problem of ``splitting'' of the images of nearby
galaxies into several objects, which leads to two interrelated
effects: underestimated luminosity of the galaxy on the one hand,
and obtaining the spectrum of the galaxy away from its physical
center on the other hand, which introduces an error into the
redshift measurements of the galaxies. Using HyperLeda data allows
one to solve the problem of bright galaxies. Accurate integral
photometry is available for most of them, and redshift data is no
less than 96\% complete for galaxies with $B\leq15\,.\!\!^{\rm
m}5$ (Falco et al., 1999). Additionally, HyperLeda contains
a large collection of data obtained in the course of mass neutral
hydrogen line sky surveys, such as ALFALFA
(Haynes et al.,
2018) and HIPASS (Meyer et al., 2004).

Before starting the selection procedure, we transformed the
inhomogeneous galaxy photometric data from different surveys to a
single $K_s$-band magnitude in the 2MASS photometric system
(Skrutskie et al., 2006). The choice of this filter is due to a
close relation of the \mbox{$K$-luminosity} of the stellar
population of galaxies with their total stellar mass. Galaxies
brighter than \mbox{$K_s<13\,.\!\!^{\rm m}5$} usually have
original flux measurements in the 2MASS survey. HyperLeda data
were therefore used without any transformation for galaxies with
known \mbox{$K$-photometry}. The overwhelming majority of our
sample galaxy data were obtained from the SDSS survey. In order to
convert the SDSS-magnitudes to the 2MASS system we used the
following expression:
\begin{equation}
(g-{K_s})_0 = 1.907 (g-r)_0 + 1.654 (r-i)_0 + 0.684
\label{eq:SDSSto2MASS}
\end{equation}
with a standard deviation $\sigma=0.126$ (Bilir et al.,
2008, equation (15)). A detailed inspection of the galaxies in
the final sample revealed the fact that the \mbox{SDSS-photometry}
for some of them, usually for chunky and extended
\mbox{LSB-galaxies},  is obviously erroneous. In these cases we
used the \mbox{$g$-, $r$- and $z$-band} photometry from DESI
Legacy Surveys (Dey et al., 2019), unifying three
projects: the Dark Energy Camera Legacy Survey (DECaLS), the
Beijing-Arizona Sky Survey (BASS), and the Mayall $z$-band Legacy
Survey (MzLS). To estimate the $K_s$-magnitude from these data, we
operated in steps. First,  Legacy photometry was converted to the
SDSS system in the assumption that \mbox{$r-i\approx i-z
\approx0\,.\!\!^{\rm m}15$,} typical for late-type galaxies at
$z=0$ (Fukugita et al., 1995; Shimasaku et al., 2001), and then
to the 2MASS system, using equation (1).
For the DECaLS survey data, we used the  B1 conversion equations
from Abbott et al. (2021), and for the BASS and MzLS
survey, those from Zou
et al. (2019). As mentioned
above, photometry for galaxies from the EGIPS catalog was carried
out in the PS1 system. EGIPS galaxy magnitudes in the SDSS system
were derived from the conversion equations of
Tonry et al. (2012) and were afterwards converted to $K_s$
by  formula (1). Note that the final sample
contains original  2MASS photometry for the vast majority of the
central EGIPS galaxies (we had to use optical photometry only for
28 out of 764 objects). In the case of candidate satellites,
\mbox{$K$-magnitudes} were estimated from the conversion equations
for 60\% of the galaxies. However, this should not lead to a
significant luminosity underestimation, since the satellite sample
is based on  dwarf galaxies, where the influence of internal
extinction is small. All photometric data were corrected for
Galactic extinction using the calibration from Schlafly and
Finkbeiner (2011).
\begin{figure} \vspace{0.2mm}
\includegraphics[width=0.49\textwidth, bb= 15 10 835 710,clip]{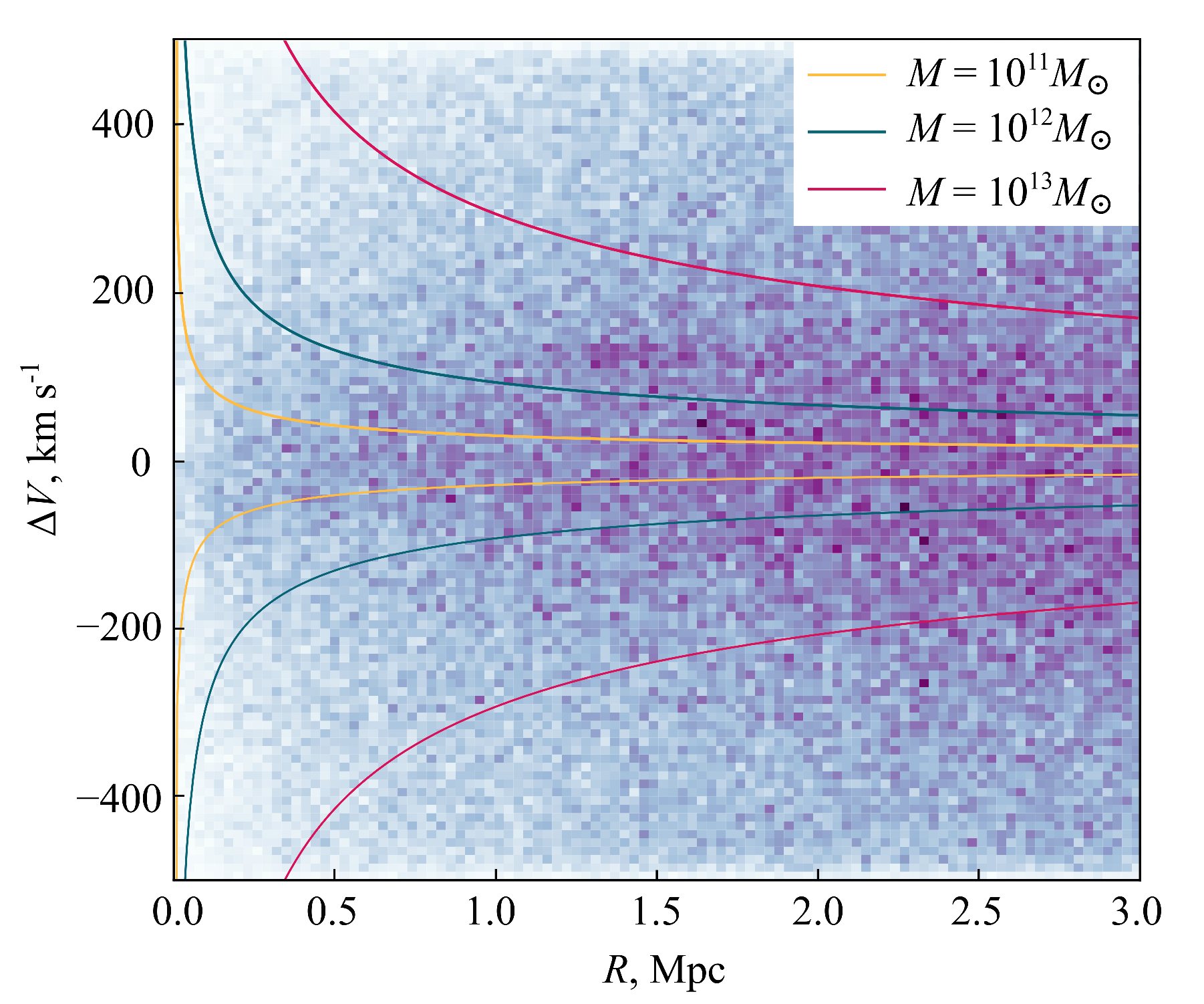}
\caption{ Distribution of neighboring galaxies in the $(\Delta
V,R)$ plane. The colored lines show the escape velocity curves for
different point masses. } \label{fig:neighbours}
\end{figure}

As the next step, a list of neighbors was compiled for each galaxy
in the EGIPS catalog. It included all galaxies within a projected
distance of less than 3~Mpc and with a velocity difference no
greater than 500~km\,s$^{-1}${} with respect to the edge-on
galaxy. The SDSS object selection algorithm often leads to
artificial splitting of extended and clumpy galaxies into a large
number of pseudo galaxies. For some of these pseudo galaxies
spectra and redshifts were obtained within the SDSS framework,
reflecting, rather, the kinematics of the galaxy itself and not
its satellites. In order to avoid the inclusion of these pseudo
galaxies in the lists of neighbors, we excluded from consideration
all objects falling into the $r$-band Petrosian ellipse of the
EGIPS galaxy   (if the $r$-band photometry was of poor quality, we
used data in other bands, primarily $g$).
Figure~1 shows the distribution of neighboring
galaxies on the \mbox{$(\Delta V,R)$} plane, where \mbox{$\Delta V =
(V_\mathrm{neighbour} - V_\mathrm{EGIPS})/(1+z)$} is the
cosmologically corrected neighbor velocity with respect to the
EGIPS galaxy, and $R$ is the projected distance between the
neighbor galaxy and the edge-on galaxy. As is evident from the
figure, the filling of the plane is approximately homogeneous at
large projected distances ($R\gtrsim 1.5$~Mpc), and a rather
pronounced concentration of neighbors is observed near $\Delta V
=0$. At this stage, we removed all  EGIPS galaxies containing a
brighter companion at a projected distance less than 1.5~Mpc and
with a radial velocity difference smaller than the escape
velocity, which corresponds to a point mass of $M=10^{13}M_\odot$.
The aim of this step is to exclude from consideration galaxies
that are part of groups with a more massive companion, and to
separate the EGIPS galaxy satellites from members of groups where
the edge-on galaxy is not the dominating galaxy in terms of mass.

To form a list of possible satellites for the considered galaxies,
we selected all objects with \linebreak
 \mbox{$|\Delta V| \leq300$~km\,s$^{-1}$} and $R<0.5$~Mpc,
 which roughly corresponds to the parameters of the known
  groups of galaxies. However, in this case we still consider
   galaxies with companions of comparable luminosity---pairs
   like the Milky Way and Andromeda. Due to projection effects,
   the task of isolating satellites of specific galaxies is also
   very complicated in such systems. We therefore excluded from the
   sample EGIPS galaxies with companions within 0.5~Mpc,
   imposing an absolute magnitude condition
\mbox{$(M^K - M^K_\mathrm{EGIPS})\leq 1^{\rm m}$}. At this stage, we
effectively selected isolated groups of galaxies where the central
galaxy, seen edge-on, dominates in luminosity over its neighbors
by a factor of at least 2.5.

\setlength{\tabcolsep}{3pt}
\renewcommand{\baselinestretch}{0.9}
\begin{table*}
\caption{Candidate satellites for EGIPS galaxies.
The following parameters are listed in the
columns: (1) and (2) give the name of the candidate
according to the HyperLeda (Makarov et al., 2014) database
and the J2000.0 coordinates of the candidate; (3) and
(4)~present the redshift of the candidate in the cosmic microwave
background reference frame in km\,s$^{-1}${} and the redshift
error; (5)--(8) give the total $g$, $r$, $i$-band magnitudes of
the candidates in the SDSS photometric system and the ${K_s}_0$
magnitudes of the 2MASS system. These magnitudes are corrected for
Galactic extinction according to Schlafly and Finkbeiner (2011); (9)
gives the absolute $K_s$ magnitude of the candidate; (10) and (11)
contain the name of the central EGIPS galaxy and its coordinates
according to Makarov et al. (2022); (12)~gives the redshift
of the central galaxy in the cosmic microwave background system;
(13) and (14)~contain the angular and projected distances between
the candidate and the central galaxy in arcminutes and kpc correspondingly;
(15)~gives the radial velocity difference corrected for
cosmological expansion
 \mbox{$\Delta V = (V_\mathrm{candidate} - V_\mathrm{EGIPS})/(1+z)$};
(16) and (17) contain the numeric values for the escape velocity
(equation~(4)) and turnaround radius
(equation~(5))  selection criteria,
correspondingly.
A sample is shown;
the full version is available as online supplementary material.} \medskip
\label{tab:Candidates}
\begin{tabular}{c|l|c|r @{.} l|.|.|.|.|.|r}
\hline
\multirow{2}{*}{No}&
\multicolumn{1}{c|}{Candidate}                             &
J2000.0                               &
\multicolumn{2}{c|}{$cz_\mathrm{CMB}$} &
\multicolumn{1}{c|}{$e_{cz}$}          &
\multicolumn{1}{c|}{$g_0$}             &
\multicolumn{1}{c|}{$r_0$}             &
\multicolumn{1}{c|}{$i_0$}             &
\multicolumn{1}{c|}{${K_s}_0$}         &
\multicolumn{1}{c}{$M_K$}   \\
\cline{2-11}
&
\multicolumn{1}{c|}{(1)}               &
\multicolumn{1}{c|}{(2)}               &
\multicolumn{2}{c|}{(3)}               &
\multicolumn{1}{c|}{(4)}               &
\multicolumn{1}{c|}{(5)}               &
\multicolumn{1}{c|}{(6)}               &
\multicolumn{1}{c|}{(7)}               &
\multicolumn{1}{c|}{(8)}               &
\multicolumn{1}{c}{(9)}               \\
\hline
 1&  SDSS\,J000043.54+143129.5    & J000043.5+143130 & 27892&2 & 10.0 & 18.69 & 17.75 & 17.75 & 15.5  & $-22.58$\\
 2&  UGC12895                     & J000038.4+200333 &  6394&4 & 2.9  & 15.86 & 15.49 & 15.49 & 14.1  & $-20.68$\\\relax
 3&  [MKB2002]J000409.64+042553.3 & J000409.6+042551 & 11488&6 & 60.0 & 18.01 & 17.66 & 17.66 & 16.28 & $-19.79$\\
 4&  AGC102304                    & J000510.6+050954 &  7852&1 & 6.2  & 17.77 & 17.43 & 17.43 & 16.11 & $-19.12$\\
 5&  PGC000425                    & J000556.1$-$135846 &  5585&2 & 70.0 &       &       &       & 14.05 & $-20.43$\\
 6&  SDSS\,J000634.50$-$004714.2    & J000634.5$-$004714 & 12677&4 & 3.7  & 19.57 & 19.28 & 19.28 & 18.1  & $-18.2\phantom{0}$\\
 7&  AGC105303                    & J000922.2+104104 &  6076&2 & 6.1  & 16.38 & 15.92 & 15.92 & 14.4  & $-20.27$ \\
 8&  PGC001031                    & J001524.4+184624 &  5215&8 & 8.5  &       &       &       & 11.75 & $-22.58$\\
 9& PGC1146389                    & J001529.0$-$001854 & 17771&7 & 2.7  & 17.2  & 16.76 & 16.76 & 15.19 & $-21.87$\\
10&  SDSS\,J001610.69$-$011244.9    & J001610.7$-$011245 & 24925&4 & 2.1  & 18.78 & 18.11 & 18.11 & 16.11 & $-21.71$\\
\hline
\multicolumn{10}{c}{}\\
\end{tabular}
\begin{tabular}{c|c | c | r @{.} l | . | . | . | . | .}
\hline
\multirow{2}{*}{No}                  &
EGIPS ID                                &
J2000.0                               &
\multicolumn{2}{c|}{$cz_\mathrm{CMB}$} &
\multicolumn{1}{c|}{$R$, arcmin}   &
\multicolumn{1}{c|}{$R$, kpc}         &
\multicolumn{1}{c|}{$\Delta V$}        &
\multicolumn{1}{c|}{$f_{V_\mathrm{esc}}$}   &
\multicolumn{1}{c}{$f_{R_\mathrm{ta}}$} \\
\cline{2-10}\
&
\multicolumn{1}{c|}{(10)}               &
\multicolumn{1}{c|}{(11)}               &
\multicolumn{2}{c|}{(12)}               &
\multicolumn{1}{c|}{(13)}               &
\multicolumn{1}{c|}{(14)}               &
\multicolumn{1}{c|}{(15)}               &
\multicolumn{1}{c|}{(16)}               &
\multicolumn{1}{c}{(17)}               \\
\hline
 1&J000042.1$+$143023 & J000042.2+143023 & 27841&1 & 1.16  & 118.0 & 46.7   & 0.15  & 0.01\\
 2&J000055.9$+$202017 & J000056.0+202017 &  6451&5 & 17.24 & 461.0 & -55.9  & 2.42  & 2.1\\
 3&J000344.7$+$041753 & J000344.8+041753 & 11205&0 & 10.08 & 449.4 & 273.4  & 26.55 & 0.92\\
 4&J000506.6$+$051213 & J000506.6+051213 &  7695&0 & 2.52  & 79.5  & 153.1  & 1.97  & 0.01\\
 5&J000556.8$-$135944 & J000556.8$-$135945 &  5411&4 & 0.99  & 22.6  & 170.8  & 1.76  & 0.0\\
 6&J000628.8$-$004702 & J000628.9$-$004703 & 12907&3 & 1.43  & 72.5  & -220.4 & 3.36  & 0.0\\
 7&J000904.2$+$105508 & J000904.2+105508 &  6324&9 & 14.74 & 387.4 & -243.5 & 25.52 & 0.83\\
 8&J001440.0$+$183455 & J001440.1+183455 &  5055&3 & 15.56 & 332.3 & 157.8  & 4.5   & 0.26\\
 9&J001539.8$-$001601 & J001539.8$-$001601 & 17613&9 & 3.96  & 267.7 & 149.0  & 8.37  & 0.35\\
10&J001615.2$-$011213 & J001615.3$-$011214 & 24839&7 & 1.26  & 115.7 & 79.2   & 0.49  & 0.01\\
\hline
\end{tabular}
\end{table*}
\renewcommand{\baselinestretch}{1}

Ideally, the used approach allows one to select isolated systems
 with a galaxy dominating in luminosity and its satellites.
 However, during visual inspection we discovered that in some
 cases the algorithm leaves for consideration galaxies that
 are probably members of clusters or groups with a neighbor
 of comparable mass. This is due to the incompleteness of redshift
 data for galaxies or to the location of the considered system at
 the periphery of a dynamically hot cluster. To clean the sample
 from such contamination, we looked over all galaxies remaining
 at this stage and excluded from consideration the ones with
 a high probability of being members of clusters or large groups. In total,
19~candidates were excluded around \mbox{14~EGIPS galaxies;}
a description of every such case is presented in Appendix~A.
We also excluded from consideration two EGIPS galaxies:
\mbox{EGIPS\,J144339.4+110821} and \mbox{EGIPS\,J131905.6$-$242504}.
In the first case, no adequate photometry is available for the
galaxy, and in the second---clear signs of interaction are
visible, distorting the velocity measurements for the galaxy.

As a result of the procedure sequence described above, we are left with
764~EGIPS galaxies and  1097~candidate satellites. The data on
these objects are presented in the Table~1 (the full version
is available online). 
\begin{figure*} \vspace{1mm}
\centering
\includegraphics[width=0.95\textwidth]{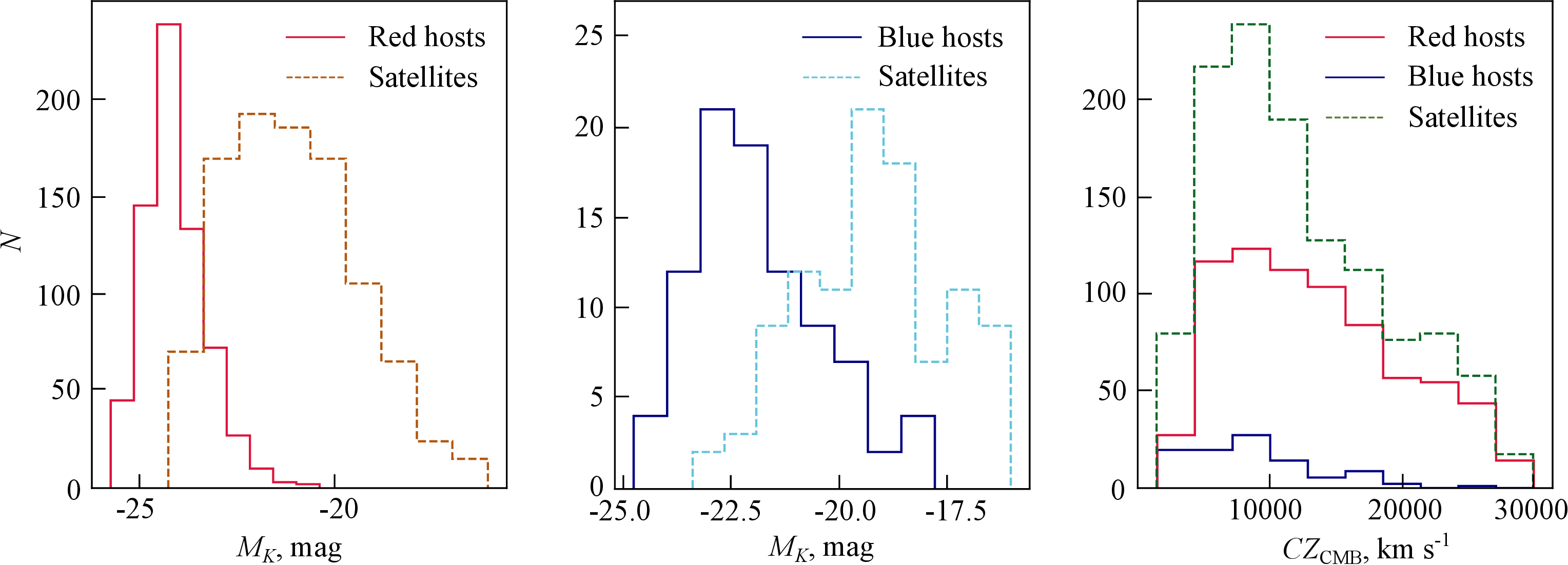}
\caption{ Histograms of the distribution of central EGIPS galaxies
and candidate satellites by absolute $K$-band magnitude ((a) for
systems with red central galaxies, (b) blue central galaxies) and
redshift (c).} \label{fig:SampleHist}
\end{figure*}

Figure~2 shows the distributions of the central
galaxies and their candidate satellites by luminosity and
redshift. The separation of central galaxies into the red and blue
ones was carried out according to the condition $(g-i)_0 \leq
0.91$, which allows one to adequately separate the galaxies of the
``red sequence'' from those of the ``blue cloud''
(Makarov et al., 2022). The distribution of central red
galaxies is characterized by a median \mbox{$K$-luminosity} of
$\widetilde{M_K}=-24\,.\!\!^{\rm m}3$, and that of the candidate
satellites---\mbox{$\widetilde{M_k}=-21\,.\!\!^{\rm m}2$}. For
central galaxies of the ``blue cloud'' the median luminosity is
equal to \mbox{$\widetilde{M_K}=-22\,.\!\!^{\rm m}1$}, and for
their satellites---\mbox{$\widetilde{M_k}=-19\,.\!\!^{\rm m}2$}.

\begin{figure*}
\includegraphics[width=0.85\textwidth]{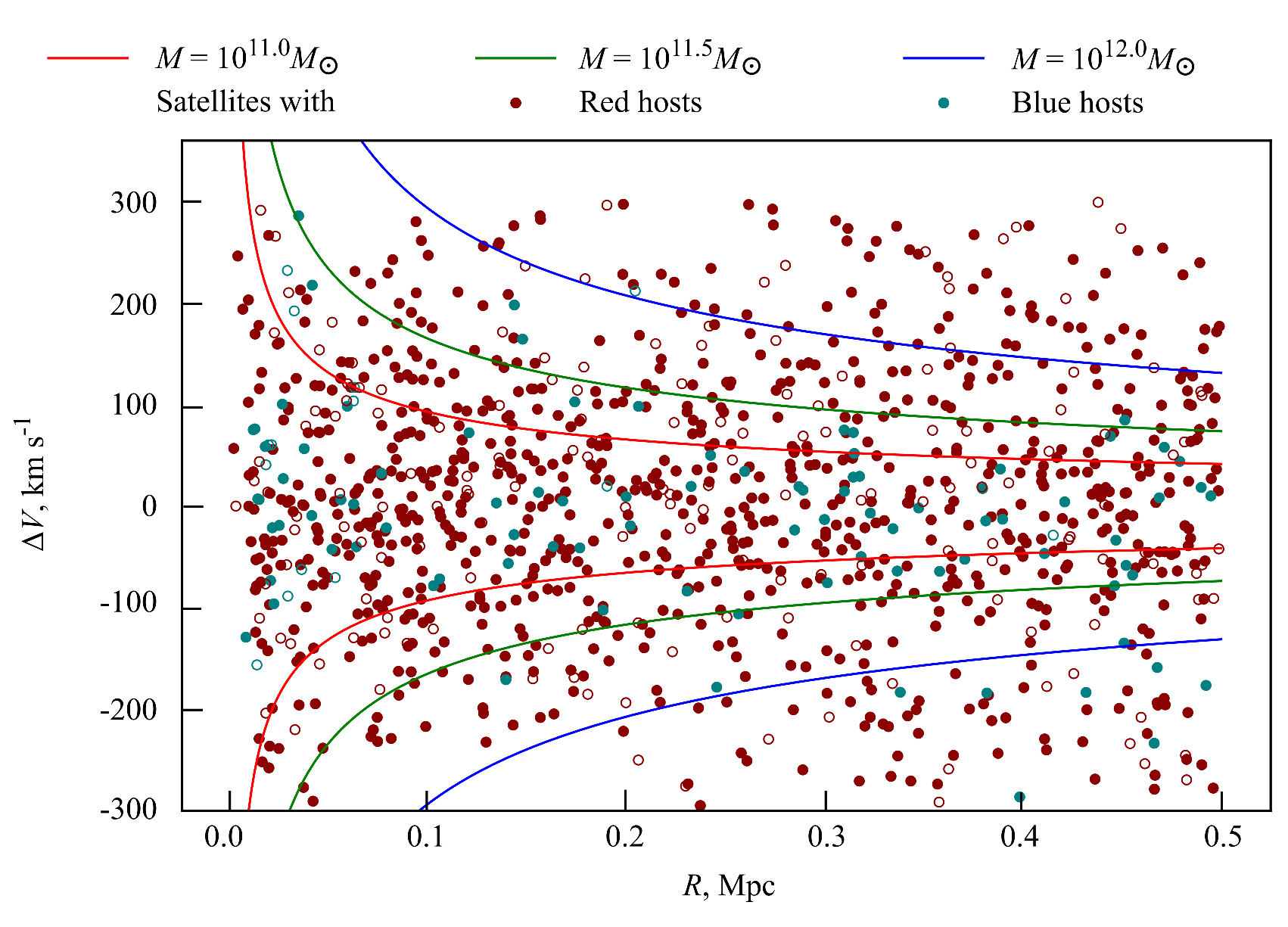}
\caption{ Distribution of candidate satellites on the $(\Delta V,
R)$ plane. The colored lines show the escape velocity curves for
different central point masses. The empty circles show galaxies
with velocity measurement errors $e_{cz}\leq 20$~km\,s$^{-1}$. }
\label{fig:dVvsR}
\end{figure*}

Figure~3 shows the distribution of all 1097~candidate
satellites for the red and blue EGIPS galaxies in the $(\Delta
V,R)$ plane. The radial velocity of most of the candidates differs
from the velocity of the central galaxy by less than
100~km\,s$^{-1}${}, which is expected for gravitationally bound
systems around isolated massive galaxies like the Milky Way.

\section{SELECTION OF GRAVITATIONALLY BOUND SATELLITES}
\label{sec:PhysicalGroups}

For a more detailed filtering of the background galaxies, we used
a technique proposed and developed in the works dedicated to the
study of galaxy groups on a scale of up to $z\sim0.01$  (Karachentsev and
Makarov, 2008; Makarov and Karachentsev, 2011, 2009).
A physical pair of galaxies should satisfy a series of restrictions.
Evidently, the total energy of a bound system should be negative,
from where it follows that the mutual velocity of the galaxies should
be lower than the escape velocity in a gravitating system,
\mbox{$ v^2 < v^2_\mathrm{esc} = 2GM/r $}. Since the observed
radial velocity difference, $\Delta V$, and the projected
distance between the galaxies,
 $R$, are obviously smaller than their spatial counterparts
 ($v$ and $r$ correspondingly), then this relation between
 the observed parameters becomes even stricter:
\begin{equation}
\Delta V^2 < v^2 < 2 \dfrac{G M}{r} < 2 \dfrac{G M}{R}.
\label{eq:Vescape}
\end{equation}
Additionally, the ``zero-velocity sphere'' radius, also called the
turnaround radius, determines the boundary between the already
collapsed high-density region and the Universe with ongoing
expansion:
\begin{equation}
r_\mathrm{ta}^3 \simeq \dfrac{ 8 G M }{ \pi^2 H_0^2 }.
\label{eq:Rta}
\end{equation}
This places a natural limit on the maximum linear size of the
system.

Assuming the relation $M=\kappa L_K$ between the total mass of a
gravitating system and its luminosity, we can rewrite the
conditions described above in terms of the  ``mass-to-luminosity''
ratio. For a physical system, the projected ``mass-to-luminosity''
ratio for the escape velocity (equation~(2)) gives
\begin{equation}
\cfrac{\Delta V^2 R}{2 G (L_{\mathrm{host}} + L_{\mathrm{sat}})} <
\kappa, \label{eq:CriterionVesc}
\end{equation}
and from the turnaround radius condition \mbox{(3)} we get
\begin{equation}
\cfrac{\pi^2\, R^3\, H_0^2}{8\, G\, (L_{\mathrm{host}} +
L_{\mathrm{sat}})} < \kappa, \label{eq:CriterionRta}
\end{equation}
where $L_{\mathrm{host}}$ and $L_{\mathrm{sat}}$ denote the
$K$-luminosity of the central galaxy and its satellite,
correspondingly. Figure~4 shows the
distributions of candidate satellites by the projected
\mbox{``mass-to-luminosity''} ratios for the escape velocity and
turnaround radius. We limited the satellite sample assuming the value
\mbox{$\kappa = 9\,M_\odot/L_\odot$} for both ratios.

As was noted by Makarov and Karachentsev (2011), the redshift
measurement errors may introduce a substantial contribution to the
velocity dispersion of galaxies in groups, especially
when these errors turn out to be comparable with the expected
velocity dispersion for the satellites, approximately or less than
100~km\,s$^{-1}${}. As is evident from Fig.~5, for
most of the candidates the radial velocity measurement error does
not exceed 20~km\,s$^{-1}$. A selection of precisely such
galaxies, therefore, became the finishing touch in our sampling of
satellites around edge-on galaxies. The final sample for analysis
of the kinematics of groups with a central luminosity dominant galaxy
 consists of  757~satellites around \mbox{547~EGIPS
galaxies}.

\section{SYSTEM MASS ESTIMATION}
The average number of candidate satellites per host galaxy in the
resulting sample does not exceed one and a half. This makes the
determination of masses of individual EGIPS galaxies very
unreliable. We assumed that the central galaxies of comparable
luminosities have roughly equal masses, and therefore, kinematics
of the surrounding satellites. Therefore, for estimating system
masses, we decided not to consider individual galaxies but
assemble groups consisting of all candidate satellites around
EGIPS galaxies of similar luminosity. We divided our sample into
10 subsamples based on the luminosity of the host EGIPS galaxy. An
approximate equality of the number of candidate satellites in each
subsample was the main criterion of the division. Unfortunately,
the luminosity range covered by the subsamples of the brightest
host galaxies turned out to be too wide. Two subsamples with a
somewhat smaller number of objects were formed from EGIPS galaxies
with $L_K\geq2.14\times10^{11}$~$L_\odot$ for further refinement.
Figure~6 shows the distribution of the candidate
satellites for each of the 10~subsamples, divided according to the
luminosity of the host  EGIPS galaxy. Each such subsample was
considered from this point on as a single composite system of
satellites.
\begin{figure*} \vspace{1mm}
\includegraphics[width=\columnwidth, bb= 1 1 769 595,clip]{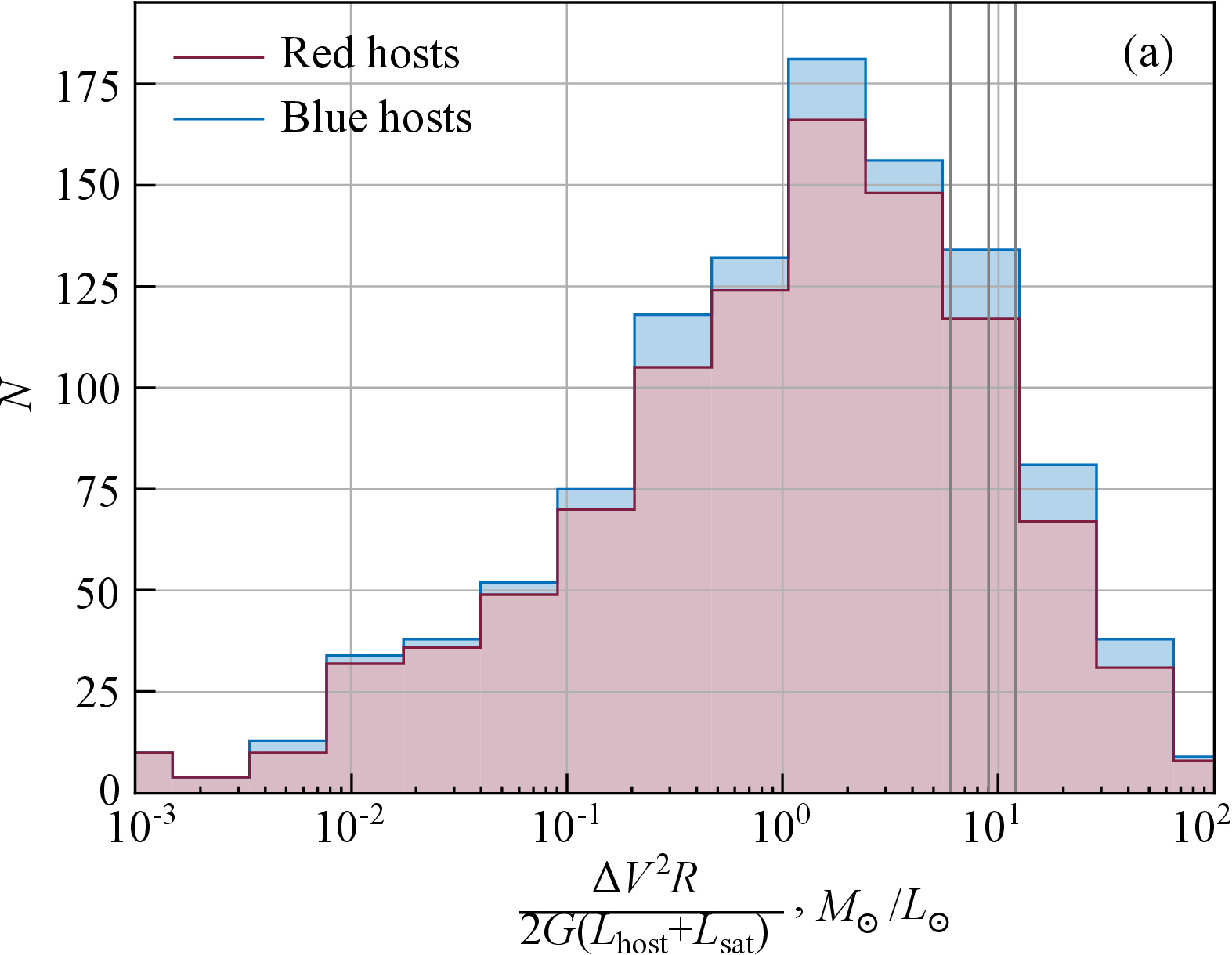} \vspace{5mm}
\includegraphics[width=\columnwidth, bb= 5 1 769 593,clip]{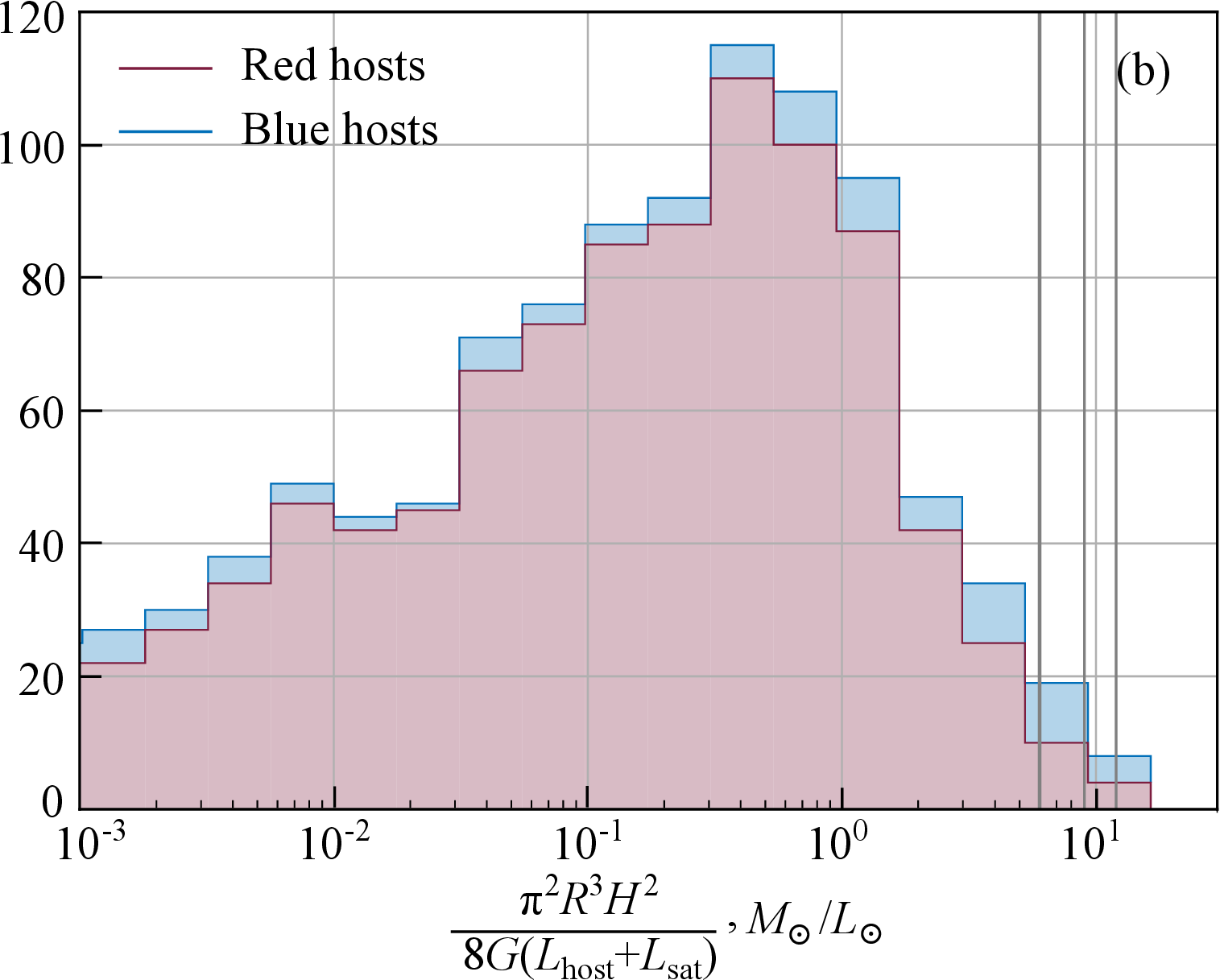}
\caption{ Cumulative histograms of the distribution of candidate
satellites by  ``virial'' (a) and  ``spatial'' parameters (b). The
vertical lines correspond to the ``mass-to-luminosity'' ratios
$\kappa=6,\,9,\,12\,{M_\odot}/{L_\odot}$. }
\label{fig:ProjectedML}
\end{figure*}

\begin{figure} 
\centering
\includegraphics[width=0.5\textwidth,bb= 0 0 750 560,clip]{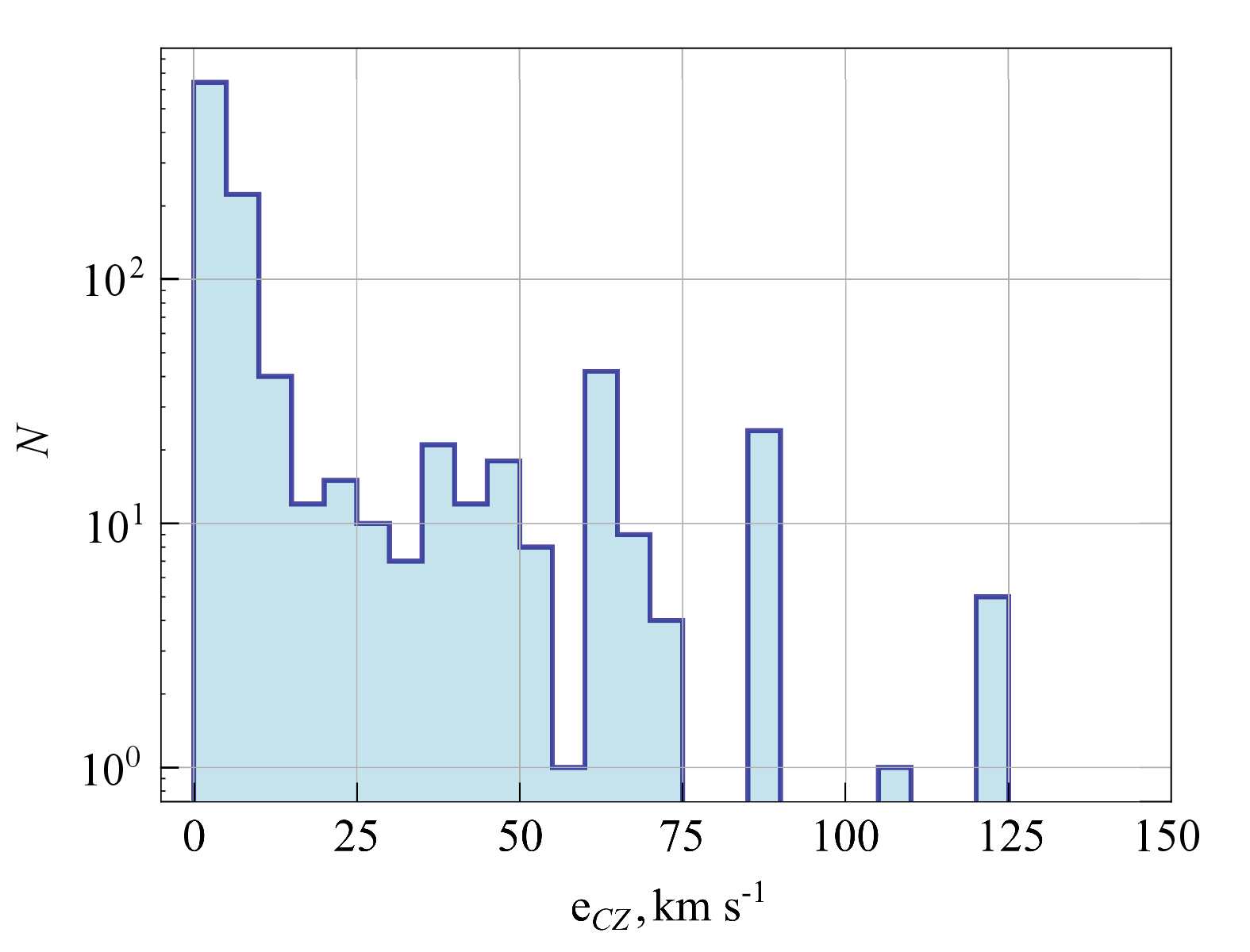}
\caption{ Distribution of candidate satellites by radial velocity
errors. The number of galaxies is presented in logarithmic scale
for better visualization of the distribution. } \label{fig:err_cz}
\end{figure}

Owing to the candidate selection algorithm for each assembled
group described above, the luminosity and mass of the satellites
turned out to be significantly lower than those of the host
galaxy. We can therefore use a simplification for estimating group
masses, which views satellites as test particles moving in the
gravitational field of the cenral galaxy.
Bahcall
and Tremaine (1981) emphasized the problems with using the
virial theorem: the mass estimate with account for projection
effects turns out to be biased, statistically ineffective and
inconsistent. They suggested an alternative to the virial
theorem---a projected mass estimator free of these shortcomings.
Unfortunately, the projected mass estimator depends on the shape of
the orbit of the moving particles. An assumption of isotropic
orbits with an average squared eccentricity equal to
 $\langle e^2 \rangle={1}/{2}$ seems most natural. In this case,
  the projected mass estimator becomes (Bahcall and Tremaine, 1981, eq.~(20))
\begin{equation}
M_p = \cfrac{16}{\pi G N} \sum_{i=1}^N \Delta V_i^2 R_i.
\label{eq:Mproj}
\end{equation}
\noindent Unlike the virial theorem, where the projection effects
make the mass estimate statistically ineffective, the dispersion
of the projected mass estimate is clearly defined and, in the case
of isotropic orbit distribution, is equal to (Bahcall and Tremaine, 1981, eq.~(22)):
\begin{equation}
\sigma^2(M_p) = \cfrac{1}{N} \left( \cfrac{128}{5\pi^2} -1 \right)
\left\langle M_p\right\rangle^2. \label{eq:SigmaMproj}
\end{equation}
\noindent We used these relations to estimate the assembled group
masses and their errors.

The average properties of the central galaxies and the
corresponding assembled groups are summarized in
Table~2. Besides the assembled groups with
designations from $L_1$ to $L_{10}$, it gives the mean parameters
for the total sample of central galaxies, as well as separately
for the  ``blue'' galaxy subsample.

As is evident from the data in the last column of Table~2 and
Fig.~6, the distribution of satellite velocities with respect
to the central galaxies turns out to be statistically unbiased.

\renewcommand{\baselinestretch}{0.9}
\setlength{\tabcolsep}{2pt}
\begin{table*}
\caption{ Main parameters of the samples under consideration. The
$M_p$ values are computed for the limit $\kappa \leq 9$ and
$e_{cz}\leq20$~km\,s$^{-1}$. The following  data are given in the
columns for each system: (1)~subsample name;  (2)---number
$N_\mathrm{host}$ of central EGIPS galaxies assembled in a group;
(3)~$K$-luminosity range covered by the given subsample;
(4)~average $K$-luminosity of the central galaxies $\left\langle K
\right\rangle_\mathrm{h}$; (5)~average $\left\langle cz
\right\rangle$ and (6)~median $\widetilde{cz}$ redshift in the
cosmic microwave background reference frame for the subsample of
central galaxies; (7)~number of satellites $N_\mathrm{sat}$ in the
assembled group; (8)~satellite velocity scattering relative to the
central galaxy, \mbox{$\sigma^2 = \sum \Delta
V_i^2/N_\mathrm{sat}$}, where the velocity difference $\Delta V$
is corrected for cosmological expansion $1/(1+z)$; (9)~harmonic
mean for the projected distances of the satellites from the host
galaxy $R_\mathrm{harm} = 1 / \left\langle 1/R \right\rangle$;
(10)~averaged total luminosity of the group with account for the
contribution from the satellites $\left\langle L_K
\right\rangle_\mathrm{g}$; (11)~estimate of the projected mass
$M_p$ of the system, derived from equation~(6);
(12)~``mass-to-luminosity'' ratio $M_p/L_K$; (13)~systemic
difference of the mean satellite system velocity with respect to
the central galaxy $\left\langle \Delta cz \right\rangle$
} 
 \medskip
\begin{tabular}{l|0|r@{--}l|.|r|r|c0|0|0|.|r@{\,$\pm$\,}l|.|r@{\,$\pm$\,}l}
\hline \multicolumn{1}{c|}{\multirow{2}{*}{Name}} &
\multicolumn{1}{c|}{\multirow{2}{*}{$N_\mathrm{host}$}} &
\multicolumn{2}{c|}{$L_K$-range,} &
\multicolumn{1}{c|}{$\left\langle L_K \right\rangle_\mathrm{h}$,}
& \multicolumn{1}{c|}{$\left\langle cz \right\rangle$,} &
\multicolumn{1}{c|}{$\widetilde{cz}$,} & &
\multicolumn{1}{c|}{\multirow{2}{*}{$N_\mathrm{sat}$}} &
\multicolumn{1}{c|}{$\sigma$,} &
\multicolumn{1}{c|}{$R_\mathrm{harm}$,} &
\multicolumn{1}{c|}{$\left\langle L_K \right\rangle_\mathrm{g}$,}
& \multicolumn{2}{c|}{$M_p$,} & \multicolumn{1}{c|}{$M_p/L_K$,} &
\multicolumn{2}{c}{$\left\langle \Delta cz \right\rangle$,}\\
& & \multicolumn{2}{c|}{$10^9 L_\odot$} &
\multicolumn{1}{c|}{$10^9 L_\odot$} &
\multicolumn{1}{c|}{km\,s$^{-1}$} &
\multicolumn{1}{c|}{km\,s$^{-1}$} & & &
\multicolumn{1}{c|}{km\,s$^{-1}$} & \multicolumn{1}{c|}{kpc} &
\multicolumn{1}{c|}{$10^9 L_\odot$} & \multicolumn{2}{c|}{$10^{11}
M_\odot$} & \multicolumn{1}{c|}{${M_\odot}/{L_\odot}$} &
\multicolumn{2}{c}{km\,s$^{-1}$}  \\
\hline \multicolumn{1}{c|}{(1)} & \multicolumn{1}{c|}{(2)} &
\multicolumn{2}{c|}{(3)} & \multicolumn{1}{c|}{(4)} &
\multicolumn{1}{c|}{(5)} & \multicolumn{1}{c|}{(6)} & &
\multicolumn{1}{c|}{(7)} & \multicolumn{1}{c|}{(8)} &
\multicolumn{1}{c|}{(9)} & \multicolumn{1}{c|}{(10)} &
\multicolumn{2}{c|}{(11)} & \multicolumn{1}{c|}{(12)} &
\multicolumn{2}{c}{(13)}\\
\hline
Full      & 547 & \multicolumn{2}{c|}{} & 114.0 & 12861 & 11605 &\phantom{0}& 757 & 103 & 84  & 132.5 & 23.2 & 1.1  & 17.5 & $  0$ & 4  \\[3pt]
$L_{1}$   &  36 &  0.28 &  22.5        &  12.1 &  7037 &  6237 &            & 38 &  47 &  33 &  13.3 &2.89 & 0.59 & 21.7 & $ -10$ & 8  \\
$L_{2}$   &  67 &  22.5 &  46.5        &  35.7 &  9061 &  8251 &            & 82 &  66 &  70 &  39.4 &  6.67 & 0.93 & 16.9 & $  0$ & 7  \\
$L_{3}$   &  76 &  46.5 &  70          &  58.6 & 10714 &  9892 &            & 91 &  90 &  77 &  66.2 & 12.3  & 1.6  & 18.6 & $  6$ & 10 \\
$L_{4}$   &  66 &  70   &  93          &  81.5 & 10554 &  9938 &            & 96 &  96 &  83 &  92.8 & 17.4  & 2.2  & 18.8 & $  3$ & 10 \\
$L_{5}$   &  71 &  93   & 113          & 103.1 & 12467 & 12238 &            & 94 & 106 & 100 & 117.8 & 20.3  & 2.6  & 17.2 & $  7$ & 11 \\
$L_{6}$   &  68 & 113   & 136          & 125.0 & 14076 & 14216 &            & 92 & 118 &  75 & 145.5 & 27.9  & 3.7  & 19.2 & $-15$ & 12 \\
$L_{7}$   &  55 & 136   & 166          & 151.6 & 14618 & 14587 &            & 88 & 114 & 107 & 179.2 & 29.8  & 4.0  & 16.6 & $  3$ & 12 \\
$L_{8}$   &  51 & 166   & 214          & 188.0 & 15976 & 17311 &            & 77 & 108 & 143 & 222.4 & 31.5  & 4.5  & 14.2 & $  9$ & 12 \\
$L_{9}$   &  38 & 214   & 300          & 246.1 & 17254 & 19521 &            & 62 & 118 & 123 & 285.3 & 35.4  & 5.7  & 12.4 & $ -7$ & 15 \\
$L_{10}$  &  19 & 300   & 440          & 346.2 & 18485 & 20118 &            & 37 & 140 & 119 & 415.3 & 64.3  & 13.3 & 15.5 & $-11$ & 23 \\
Blue      & 48  & \multicolumn{2}{c|}{} & 29.7 & 10120 &  9289 &            & 51 &  64 &  52 & 34.0  &  7.6  & 1.3  & 22.5 & $ -9$ & 9  \\ 
\hline
\end{tabular}
\label{tab:Groups}
\end{table*} \renewcommand{\baselinestretch}{1.0}

There is a close correlation between the total luminosity of an
assembled group and its projected mass, shown in
Fig.~7. The interrelation between mass and
$K$-luminosity is well-described by a linear dependence in
logarithmic scales
\begin{equation}
\log M_p = (0.875\pm0.028)\log \left\langle L_K
\right\rangle_\mathrm{g} + (2.59\pm0.29). \label{eq:MvsL}
\end{equation}
\noindent The slope of the dependence at a $4\,\sigma$ level is
less than one.

\section{CONCLUDING STATEMENTS}
\begin{figure*} \vspace{1.5mm}
\centering
\includegraphics[width=0.85\textwidth]{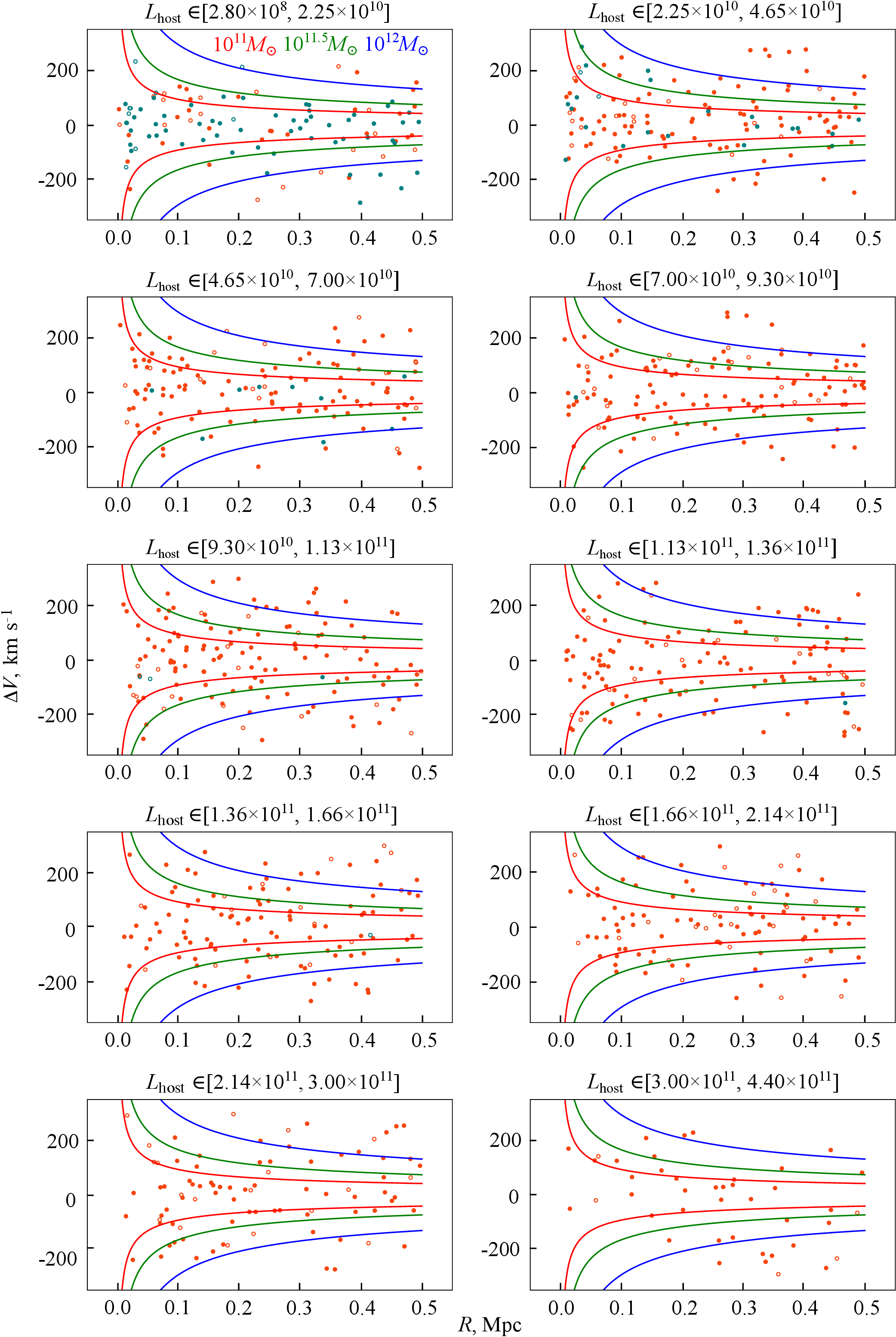}
\caption{ Distribution of candidate satellites in the $(\Delta V,
R)$ plane, divided into subsamples according to the luminosity of
the host galaxy. The designations are the same as in
Fig.~3. } \label{fig:dVvsRvsLum}
\end{figure*}

\begin{figure} \vspace{1mm}
\includegraphics[width=.47\textwidth]{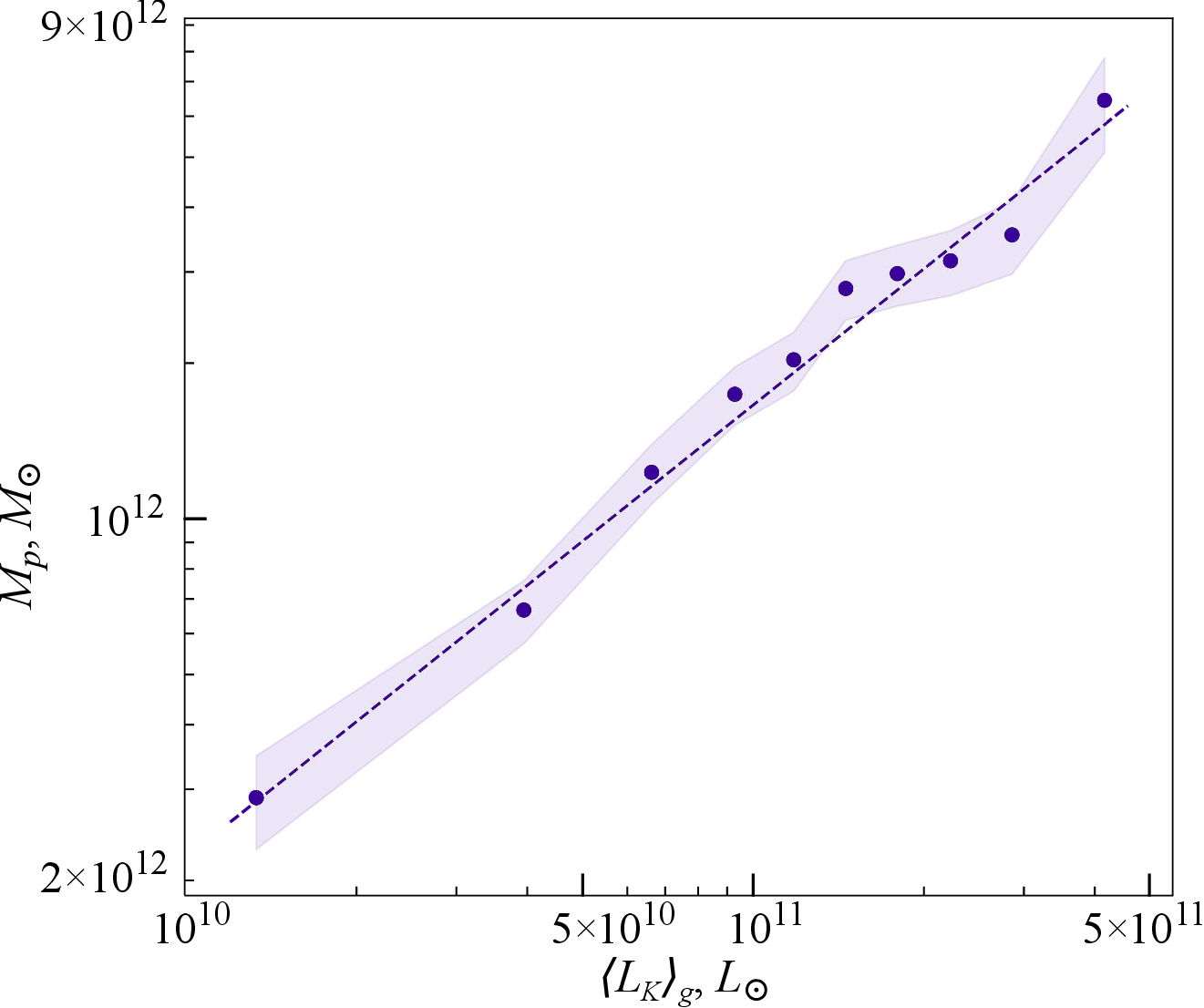}
\caption{ The dependence of assembled group mass on luminosity.
The linear regression from equation~(8) is shown by
the dashed line. The shaded regions mark the  $1\sigma$ interval
of the mass estimate errors. } \label{fig:MvsL}
\end{figure}

Using photometry and redshifts obtained in the  SDSS\,DR17
(Abdurro'uf et al., 2022) survey and
collected in the HyperLeda (Makarov et al., 2014) database, we carried out a search for
possible satellite candidates around edge-on galaxies fron the
EGIPS (Makarov et al., 2022) catalog. We searched for systems
with a central luminosity dominant galaxy: $(M^K -
M^K_\mathrm{EGIPS}) > 1^{\rm m}$. We found 1097~candidate
satellites around \mbox{764~EGIPS galaxies} with radial velocities
in the range of  $cz_\mathrm{LG}\geq 2\,000$ and
\mbox{$cz_\mathrm{CMD}\leq30\,000$~km\,s$^{-1}$} with projected
distances of less than 500~kpc and radial velocities of less than
300~km\,s$^{-1}${} with respect to the central galaxy. Out of
these, 757~satellites around 547~central galaxies have redshift
 accuracies better than 20\,km\,s$^{-1}$ and satisfy
the criterion of a gravitationally bound system, which makes them
suitable for determining the physical parameters of the systems.
The population of satellites has a typical projected distance of
84~kpc and a mean velocity dispersion of 103~km\,s$^{-1}${}. The
distribution of satellite velocities relative to the central
galaxies is statistically unbiased.
Assuming an isotropic nature of the satellite orbits with an
average eccentricity of \mbox{$\langle e^2 \rangle = {1}/{2}$}, we
estimated the total (projected) masses of EGIPS galaxies. In a
wide interval of luminosities the projected mass estimates for the
EGIPS galaxies follow the dependence \mbox{$M_p \propto
\left\langle L_K \right\rangle_\mathrm{g}^{0.88}$} with an average
ratio $(17.5 \pm 0.8)\,M_\odot/L_\odot$, typical for nearby
massive spiral galaxies (Karachentsev and Kashibadze,
2021; Karachentsev and Kudrya, 2014). Numerous works dedicated to studying
 galaxy groups have revealed that the \mbox{``mass-to-luminosity''}
ratio of the systems increases with luminosity and mass. Thus,
Makarov and Karachentsev (2011) used a sample of approximately
11~thousand galaxies on the scale of the Local Supercluster
\mbox{$cz_\mathrm{LG}\leq 3500$~km\,s$^{-1}$} to find the
dependence $M \propto L_K^{1.15}$. In that work, the overwhelming
majority are groups with luminosity $L_K \gtrsim
10^{10}\,L_\odot$ and a median $L_K=1.2 \times 10^{11}\,L_\odot$.
However, when switching to systems of lower luminosities, the
behavior of the dependence changes dramatically. In their studies
of dwarf galaxies, Tully (2005) and Tully et al. (2006) discovered
that low luminosity groups
demonstrate a significantly higher ``mass-to-luminosity'' ratio
compared to the ``normal'' groups. A similar result was obtained
by  Makarov and Uklein (2012). Kourkchi and Tully
(2017) postulated the dependence \mbox{$M/L_K \propto L_K^{-0.5}$} for
systems with luminosities \mbox{$L_K<9.27 \times 10^8\,L_\odot$}.
In their studies of the nearby groups of galaxies at distances
less than 11~Mpc in the Local Volume, probably the most
studied region of the Universe,
 Karachentsev and Kashibadze (2021) have shown that low luminosity groups
  are well-described by this dependence. The transition between
  the dependences for groups of low and high luminosity takes
  place at \mbox{$L_K\simeq 3\times10^{10}\,L_\odot$}.
The \mbox{``mass-to-luminosity''} ratio near this luminosity
reaches its maximum value, which indicates the most effective
transformation of matter into stars for systems of this
luminosity. The central dominating galaxies of the groups in our
work occupy precisely the luminosity range falling into the
transition zone. Expression~(8) and
Table~2 demonstrate a gradual growth of the
\mbox{``mass-to-luminosity''} ratio into the region of low
luminosity galaxies.

As the next stage of our work we propose investigating the
orientation and nature of the satellite motion with account for
the direction of rotation of the  EGIPS galaxies themselves. This
task became relevant lately due to the discovery of planar
structures of the satellites around the Milky Way, Andromeda, and
other massive nearby galaxies (Kroupa et al., 2005;
Koch and Grebel, 2006; Ibata et al., 2013; Pawlowski et al., 2015; Tully et al., 2015; Libeskind
et al., 2019; Mart{\'\i}nez-Delgado et al., 2021),
being a serious challenge for the standard cosmological model on
the scale of the order of 100~kpc.


\begin{acknowledgments}
We are grateful to the referee for valuable comments that helped
us significantly improve the paper. We acknowledge the usage of the
HyperLeda database ({\url{https://leda.univ-lyon1.fr/}}).
Funding for the Sloan Digital Sky Survey~V has been provided by
the Alfred~P.~Sloan Foundation, the Heising-Simons Foundation, the
National Science Foundation, and the Participating Institutions.
SDSS acknowledges support and resources from the Center for
High-Performance Computing at the University of Utah. The SDSS web
site is \url{www.sdss.org}. SDSS is managed by the Astrophysical
Research Consortium for the Participating Institutions of the SDSS
Collaboration, including the Carnegie Institution for Science,
Chilean National Time Allocation Committee (CNTAC) ratified
researchers, the Gotham Participation Group, Harvard University,
Heidelberg University, The Johns Hopkins University, L'Ecole
polytechnique f\'{e}d\'{e}rale de Lausanne (EPFL),
Leibniz-Institut f\"{u}r Astrophysik Potsdam (AIP),
Max-Planck-Institut f\"{u}r Astronomie (MPIA Heidelberg),
Max-Planck-Institut f\"{u}r Extraterrestrische Physik (MPE),
Nanjing University, National Astronomical Observatories of China
(NAOC), New Mexico State University, The Ohio State University,
Pennsylvania State University, Smithsonian Astrophysical
Observatory, Space Telescope Science Institute (STScI), the
Stellar Astrophysics Participation Group, Universidad Nacional
Aut\'{o}noma de M\'{e}xico, University of Arizona, University of
Colorado Boulder, University of Illinois at Urbana-Champaign,
University of Toronto, University of Utah, University of Virginia,
Yale University, and Yunnan University.

\end{acknowledgments} \vspace{-2mm}
\section*{FUNDING}
This work was carried out with the financial support of the RSF
grant \mbox{No.~19-12-00145}. \vspace{-2mm}

\section*{CONFLICT OF INTEREST} The authors declare no conflict of
interest regarding the publication of this paper.  \vspace{-2mm}

\section*{APPENDIX~A}
\section*{GALAXIES EXCLUDED FROM CONSIDERATION}\label{sec:cluster_members}

Information about the membership of galaxies in clusters and their
radial velocities was taken from the
HyperLeda (Makarov et al., 2014) database.

\begin{list}{$\bullet$}{
\setlength\leftmargin{4.5mm} \setlength\topsep{1.4mm}
\setlength\parsep{0mm} \setlength\itemsep{1mm} } \item
SDSS\,J000301.88+331036.1, a candidate satelli\-te of
\mbox{EGIPS\,J000312.06+331118.0}, projects close to the tight
pair \mbox{SDSS\,J000302.29+331032.6} and
\mbox{SDSS\,J000302.15+331024.6}, which are part of a galaxy
cluster. With the exception of the  {EGIPS galaxy} and the
candidate satellite, redshift measurements are unavailable for the
remaining cluster members; however, evidently, the cluster is real
and these two galaxies are its members.

\item PGC\,1157067 and its main galaxy \linebreak
\mbox{EGIPS\,J011551.3+000848} are located at the periphery of
the  ABELL\,168 
cluster. The radial velocity relative to the cluster center
is approximately 1700~km\,s$^{-1}${}. 

\item EGIPS\,J030718.3$-$093645 (NGC\,1216) and its candidate
satellite \mbox{HCG\,023:[dRC97]26} are, probably, members of a
compact HCG023 (Hickson, 1982) group. Their radial
velocities differ from the group velocity by  775 and
685~km\,s$^{-1}${} correspondingly.

\item The EGIPS\,J031922.9+004924 galaxy and its candidate satellites
\mbox{SDSS\,J031844.25+005302.5} \linebreak  and
\mbox{PGC\,1176312} are excluded from consideration due to a
possible connection with the \linebreak
\mbox{[BFW2006]\,MR18\_06184} group with a radial velocity of
10\,949~km\,s$^{-1}${}.

\item As is evident in the image, the 
EGIPS\,J074900.3+\linebreak370942 galaxy
(with PGC\,3128739 as its
candidate satellite) is interacting with a close companion,
PGC\,021876, which has no redshift data. The magnitude difference
is less than $1^{\rm m}$, so the system should not have passed the
second stage of selection.

\item EGIPS\,J075939.3+263309 and its candidate satellite
\mbox{PGC\,1781379} are located at the periphery of the IC\,486,
IC\,485, IC\,484 group. We excluded it from consideration to avoid
the edge effects, despite the rather large velocity difference of
672~km\,s$^{-1}${}.

\item The  EGIPS\,J080613.4+174223 (NGC\,2522) 
with all its
candidates satellites---\mbox{PGC\,1534373}, \linebreak
PGC\,1536557, \mbox{SDSS\,J080627.26+174317.1} and
PGC\,1548640---are members of the USGC\,U167
 (Ramella et al., 2002) group. It was missed by the
 selection algorithm due to the velocity difference between
\mbox{EGIPS\,J080613.4+174223} and the other known companions of
the USGC\,U167 group of over 300~km\,s$^{-1}${}.

\item A sparse structure of galaxies is observed in the direction
of EGIPS\,J104046.9$-$090040 and its candidate satellite
\mbox{PGC\,996238} with a radial velocity of  9000~km\,s$^{-1}${},
close to the velocity of the pair of galaxies under consideration.
The ABELL\,1069 cluster is located in the background.

\item The image shows that 
\mbox{EGIPS\,J155604.4+402747} (its candidate satellite is
SDSS\,J155543.93+\linebreak402703.8) is interacting with a close
brighter companion,  \mbox{SDSS\,J155605.25+402752.8}, with no
redshift measurements available. As a consequence, the pair was
missed by the algorithm.

\item
The central object EGIPS\,J160459.5+235812, with PGC\,2567181 as a
candidate satellite, is a member of the AWM\,4
(Koranyi and Geller, 2002) cluster. The radial velocity
difference of 1719~km\,s$^{-1}${} turned out to be too large for
an association of the \mbox{EGIS galaxy} with the cluster by our
algorithm.

\item \mbox{EGIPS\,J161234.0+253002} (its candidate  satel\-lite is
\mbox{SDSS\,J161155.08+253725.8)} possibly forms a group with
\mbox{SDSS\,J161212.79+253340.3}, which is a spiral galaxy  of comparable
luminosity. Only the photometric redshift is known for the latter,
$z_\mathrm{photo}=0.033$ (Bilicki et al., 2014), which does
not contradict the suggestion that they form a bound system with
the \mbox{EGIPS\,J161234.0+253002} galaxy.

\item EGIPS\,J171524.2+572101 (NGC\,6345) and its candidate
satellite \mbox{PGC\,2567181} were excluded due to a possible
connection to the group of galaxies around NGC\,6338, despite the
1737~km\,s$^{-1}${} radial velocity difference.

\item
An \mbox{XLSB galaxy} PGC\,3441769 of comparable size is projected
onto the central galaxy  \linebreak
\mbox{EGIPS\,J225650.8$-$085803}   (its candidate satellite is
PGC070056). The galaxies likely form a close pair.

\item \mbox{EGIPS\,J234800.6+272231} together with its candidate
satellites \mbox{PGC\,1809943} and \mbox{PGC\,085775} are located
at the periphery of the \mbox{ABELL\,2666} cluster. The radial
velocity relative to the cluster center is $~$480~km\,s$^{-1}${}.
\end{list}

\clearpage
\end{document}